\documentclass[review]{elsarticle}
\usepackage{epsfig,journals}

\usepackage{lineno,hyperref}

\journal{Journal of \LaTeX\ Templates}

\biboptions{authoryear}

\def\citen#1{\citeauthor{#1} \citeyear{#1}}

\usepackage{color}

\newcommand{\scri}{\scriptsize}
\newcommand{\refresp}{}
\newcommand{\refrespb}{}
\newcommand{\refrespc}{}

\begin{document}

\begin{frontmatter}

\title{Precursors of an upcoming solar cycle at high latitudes from
coronal green line data}

\author{K. Petrovay}
\cortext[mycorrespondingauthor]{Corresponding author}
\ead{K.Petrovay@astro.elte.hu}
\author{M. Nagy}
\author{T. Gerj\'ak}
\address{E\"otv\"osUniversity, Department of Astronomy, Budapest, Hungary}
\author{L. Juh\'asz}
\address{E\"otv\"osUniversity, Department of Geophysics and Space Science,
Budapest, Hungary}

\begin{abstract}
After reviewing potential early indicators of an upcoming solar cycle at high
latitudes, we focus attention on the rush--to-the-poles (RTTP) phenomenon in
coronal green line emission. Considering various correlations between properties
of the RTTP with the upcoming solar cycle we find a correlation between the rate
of the RTTP and the time delay until the maximum of the next solar cycle. On the
basis of this correlation and the known internal regularities of the sunspot
number series we predict that, following a minimum in 2019, cycle 25 will peak
in late 2024 at an amplitude of about 130 (in terms of smoothed monthly revised sunspot
numbers). This slightly exceeds the amplitude of cycle 24 but it would still
make cycle 25 a fairly weak cycle.
\end{abstract}


\end{frontmatter}


\section{Introduction}

Solar Cycle 25 reached its maximum in April 2014 with a 13-month
smoothed sunspot number of $R=116$ (revised value\footnote{\refrespb The
revision of the official sunspot number series that took place in 2015
was a well known crucial milestone in recent solar physics. For more 
explanations see {\tt
http://sidc.oma.be/silso/newdataset}}) or $R=70$ (unrevised value),
about $30\,$\% lower than an average cycle and roughly half the
typical amplitude of cycles 18--23, comprising the Modern Maximum. As
common with solar cycles, the current cycle was double peaked: the
main maximum was preceded by a lower secondary maximum in March 2012.
The two maxima were associated with a phase shift between the
hemispheres, the first and second maxima corresponding to peak
activity in the Northern and Southern hemisphere, in September 2011
and April 2014, respectively.

The evident end of the Modern Maximum has triggered increased interest in the
underlying causes of intercycle variations in solar activity and in forecasting
methods (\citen{Petrovay:LRSP}). However, despite extensive efforts, attempts to forecast an upcoming
solar cycle significantly earlier than its start have not been successful.
The most successful methods of solar cycle prediction are either based on the
Waldmeier effect (relation between the rise rate of activity in the early phase
of a solar cycle to its maximum), for which the cycle must have already started,
or on the polar precursor method (a link between indicators of the magnetic
field at the poles of the Sun, peaking around solar minimum, and the amplitude
of the next maximum).

Both of these methods suggest that the most promising place on the Sun to look
for even earlier precursors of an upcoming cycle is at high heliographic
latitudes. On the one hand, solar cycles are known to overlap by up to $\sim
1$--$2$ years on the sunspot butterfly diagram and possibly up to $\sim 5$--$6$
years on the butterfly diagrams of other solar features such as torsional
oscillations, ephemeral active regions or coronal line emission. On the other
hand, the polar magnetic field is observed to be built up from the poleward
transport of following ($f$) polarity magnetic flux from decaying active regions
by meridional flows and turbulent diffusion, traced by a high-latitude poleward
branch in the butterfly diagram of various solar features such as polar faculae
and coronal line emission. Detailed studies of the evolution of the latitudinal
distribution of these proxies may therefore potentially yield some early signs
of the new solar cycle in the making.

In the present paper, after briefly reviewing information about some other
potential proxies, we will focus on coronal green line emission data, in
particular to the poleward branch on the butterfly diagram of this emission,
also known as the ``rush to the poles'' (RTTP). As the green line emission in
this branch is thought to trace the closed magnetic field structures just
outside the boundary of the polar coronal hole, the RTTP is a manifestation of
the decrease of poloidal magnetic flux due to the poleward spread of opposite
polarity flux from the $f$-polarity parts of active regions of the ongoing
cycle. Indeed, soon after the RTTP finally reaches the pole the polar magnetic
field is reversed, and, as a consequence of further flux transport from the
lower latitudes, starts to increase in amplitude again until it reaches its
maximum around the following sunspot minimum. This raises the question whether
properties of the RTTP can give us a hint regarding the following increase of
the polar field to its maximum, and, by implication, regarding the timing and
amplitude of the next sunspot maximum.

Section 2 reviews information about other high-latitude magnetic proxies in view
of forecasting possibilities for cycle 25. Section 3 presents the green line
data used, our methods of processing these data, and  the results, while Section
4 concludes the paper.

\section{The high-latitude magnetic field and its proxies}

In this section we review information about other high-latitude magnetic proxies
in view of forecasting possibilities for cycle 25. For further and less
cycle-specific information we refer the reader to the reviews of
\cite{PetrieLRSP} and \cite{Petrie+}.

\begin{figure}
\epsfig{figure=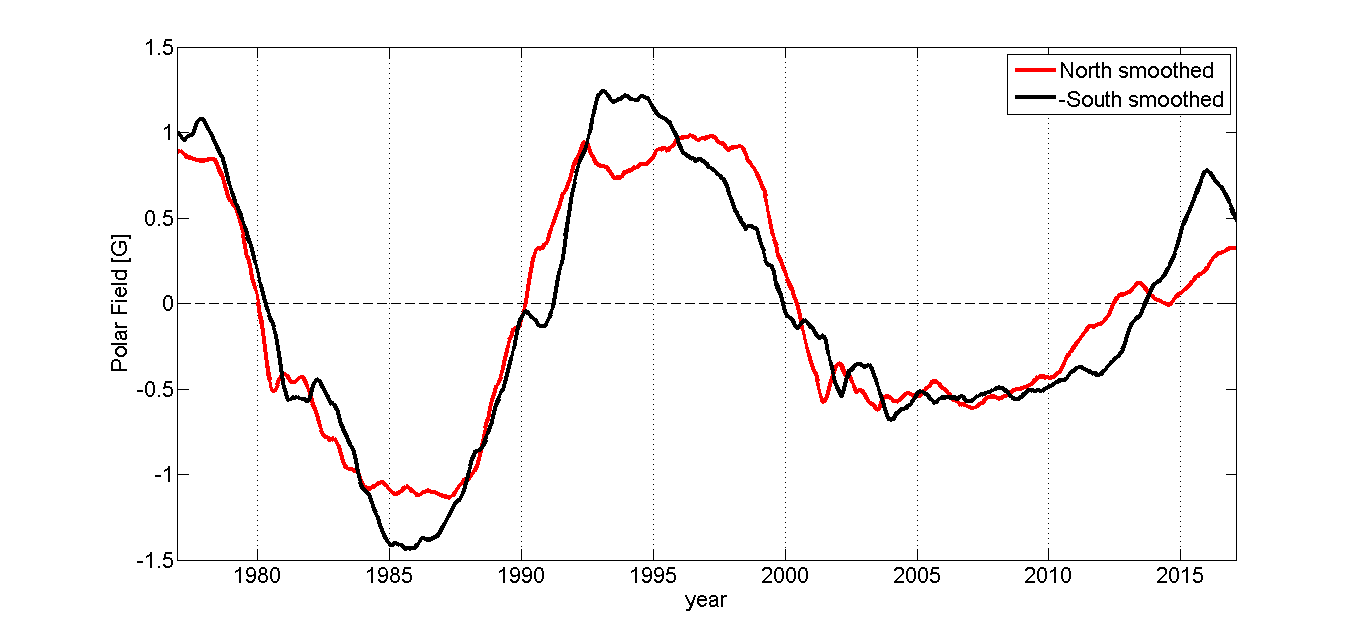,width=\textwidth}
\caption{WSO polar magnetic field strength smoothed with a 13-month
sliding window for the N (red)  and S (black) hemispheres.}
\end{figure}

\subsection{Polar magnetic fields}

Direct measurements of the magnetic field strength near the Sun's poles have
been made on a daily basis at Wilcox Solar Observatory since 1976. What is
measured is the mean value of the line-of-sight field component in the polemost
aperture of the instrument, covering roughly the area polewards of $55^\circ$
latitude. Upon passing data through a low-pass window to filter out annual
variations due to the tilt of the Sun's axis and averaging between hemispheres,
the resulting polar field amplitude is found to have peaked in the range
0.65--1.31 G for the last four solar minima. The WSO polar field strength last
reversed in March 2013 (one year after the peak of sunspot acitvity in the N
hemisphere and one year ahead of the peak in the S hemisphere). Considering the
two hemispheres separately, the reversal in the Northern hemishphere took place
in May 2012 while in the South it occurred in July 2013 (Fig.~1
{\refrespb and
Table~1}). In
the 4 years elapsed since the reversal, the polar field reached a peak value of
0.54 G in February 2016, significantly below even the lowest previous peak of
0.65 G in the last solar minimum. While, based on previous experience, the
current decrease of the polar field may still be reversed and a higher maximum
value may be reached, in none of the three previous solar minima did the peak
WSO polar field strength exceed the highest value measured in the first 4 years
after reversal by more than $\sim 20\,\%$. This suggests that the peak polar
field during the upcoming solar minimum may not exceed the value of 0.65
measured in the last minimum. As the peak polar magnetic field has been shown to
correlate with the amplitude of the following sunspot maximum, by inference we
may conclude that cycle 25 is unlikely to be significantly stronger than cycle
24. Instead it may peak at amplitude similar to or slightly lower than the
current cycle.

It should be noted that the reversal of the polar field measured at WSO is not,
strictly speaking, the actual polar reversal. Even in the case of an axially
symmetric magnetic field distribution, the mean field poleward of $55^\circ$
latitude reverses when the magnetic neutral line is still at a considerable
distance from the pole. An alternative method to define the moment of field
reversal is to consider the moment when the axial dipole moment (as determined
from a spherical harmonic decomposition) of the global solar magnetic field
distribution changes sign. On the basis of their analysis of SDO HMI
magnetograms, \cite{Sun2015} report that this took place in October 2013; the
mean field above $60^\circ$ latitude reversed in November 2012 and March 2014 on
the N and S hemispheres, respectively.

H$_\alpha$ filaments and filament channels serve as a good proxy tracing the
position of the neutral line. Filament observations at Kislovodsk indicate that
the neutral line finally reached the N pole in July 2013 and the S pole in
December 2014.

Magnetic measurements at WSO and with the newly installed 
{\refrespc Routine Prediction Solar Telescope (RPST)}
at Kislovodsk have shown an interesting phenomenon: while the magnetic field
strength in the South polar region has been increasing at a steady rate since
the polar reversal, in the North polar region it has lingered at very low
values in the first six months following reversal before starting to increase at
a slower rate compared to the South. As a result, the magnetic field strength at
N pole is still significantly weaker compared to the South, a situation somewhat
analoguous to the minimum prior to cycle 23. Nevertheless, in cycle 23 the
activity level in the two hemispheres did not prove to be very different and it
actually peaked two years earlier in the North.

\begin{table}
\caption{Milestones of solar cycle 24}
\begin{minipage}{\textwidth}
\begin{tabular}{lrrrl}
\hline
Latitude & North & South & Overall \\
\hline
Sunspot maximum & 2011.7 & 2014.3 & 2014.3 & \cite{SILSO}\\
Reversal $>55^\circ$ lat. (WSO) & 2012.5 & 2013.6 & 2013.3 & \cite{WSO}\\
Reversal $>60^\circ$ lat. (HMI) & 2012.9 & 2014.3 & 2013.8\footnote{global
dipole} & \cite{Sun2015}\\
Reversal at $90^\circ$ lat. (Kislovodsk) & 2013.6 & 2014.9 & & \cite{Tlatov2015}\\
\hline
\end{tabular}
\end{minipage}
\end{table}

\subsection{Torsional oscillations}

The realization that the equatorward migration of a pair of belts of alternating
fast and slow rotation (known as torsional oscillation) starts at latitudes
$\sim 50^\circ$--$60^\circ$ many years before the appearance of the first
sunspots in the region of maximal shear was what originally gave rise to the
concept of an extended solar cycle in the 1980s (\cite{Wilson1988}). 
{\refresp
In this scenario the high-latitude component of the torsional oscillation
pattern is interpreted as the extension of the low latitude pattern backward in
time, implying that solar cycles are considerably longer than 11 years,with a
long overlap between consecutive cycles.} Indeed,
the lower latitiude, fast-rotating member of the new pair appeared roughly 10
years before the start of cycle 24. A few years ago considerable alarm was
caused by the claim that no sign of the new belt pair 
{\refresp that, in the extended solar cycle scenario, is taken to correspond}
to cycle 25
could be seen at high latitudes, despite being long overdue ---some even
envisioned the onset of a new grand minimum. It was, however, demonstrated by
\cite{Howe2013} that if the reference background differential rotation profile
used in the derivation of the rotational modulation is a mean calculated over a
shorter time period, the torsional oscillation signal associated with cycle 25
is clearly seen starting from $\sim 2010$. At any rate, the detected signal is
significantly weaker than it was in the case of cycle 24, possibly
indicating a continued long term weakening of the solar activity level.
Nevertheless, as helioseismic {\refresp and surface Doppler} 
data on the high latitude torsional oscillations {\refresp (\citen{Ulrich1988},
\citen{Howe2006})}
have only been available for a few solar cycles 
{\refresp (cf. Fig. 30 in \citen{Howe2009})}, 
far reaching conclusions
regarding a relationship between this pattern and cycle amplitudes should be
treated with caution.

\subsection{Polar faculae}

Polar faculae are generally regarded to be a manifestation of the intermittent
nature of the photospheric magnetic field at high latitudes, i.e. the equivalent
of magnetic network elements at lower latitudes, with higher typical magnetic
flux as a consequence of the higher $\sim 10$ G strength of the large scale
field near the poles. This interpretation was recently empirically confirmed by
\cite{Kaithakkal2013} on the basis of Hinode observations: nearly all magnetic
patches with a flux above $10^{18}$ Mx were found to be unipolar and to harbour
polar faculae. A few faculae  were found to correspond to minority polarity
patches but these tend to have fluxes below $10^{18}$ Mx. (\citen{Blanco2010}
report that 15\,\% of polar faculae possess minority polarity magnetic field.)

All this implies that counts of polar faculae can be used as proxies of the
polar magnetic field strength. Indeed, \cite{Munozjara2012} report an excellent
correlation between Mount Wilson Observatory polar facular counts (above
latitudes of $70^\circ$) and WSO magnetic field strength measurements,
effectively extending the polar magnetic field data set back to 1906. For a
similar reconstruction based on Kodaikanal Ca II K data see \cite{Priyal2014}.

\subsection{Ephemeral active regions}

Active regions tend to emerge at the latitude of maximal shear between the pair
of fast and slow rotating belts of the torsional oscillation pair. This suggests
that the pattern is either due to a magnetic quenching of the poleward angular
momentum transport in the shallow layers of the convective zone
(\citen{PetrovayFDE2002}) or, indirectly, is a consequence of the Coriolis force
acting on meridional inflows towards the activity belt caused by a magnetic
suppression of the convective heat transport (\citen{Spruit}). Either way, one
may expect that the backwards extension of the equatorward migrating torsional
oscillation belt is also accompanied by the emergence of bipolar magnetic
regions around the region of maximal shear. Before the appearance of the first
sunspots of a cycle these regions can only be detected as ephemeral active
regions (EARs). It was indeed found long ago by \cite{Martin+Harvey79} that high
latitude ephemeral active regions tend to cluster on the backwards extension of
the wings of the sunspot butterfly diagram (cf. also \citen{Hagenaar2001}).

More recently, in their analysis of Kitt Peak and MDI magnetograms for three
solar cycles \cite{Tlatov2010} find that the first bipoles with orientation
corresponding to an upcoming solar cycle appear at latitudes of $\sim 60^\circ$,
shortly after the maximum of the previous cycle.

{\refresp 
 \cite{McIntosh2014} and \cite{McIntosh2017} study the migration pattern of EUV
coronal bright points and of certain magnetogram features that they call g-nodes
and consider to be tracers of bright points. In agreement with the extended
cycle paradigm they find that the equatorward migration of these features starts
at $55^{\circ}$ latitude many years before the appearance of the first sunspots
of the subsequent cycle. By construction, their g-nodes are unipolar flux
concentrations, so the relationship with EARs is not obvious. 
}

By an ingenious method based on the proper motions of coronal jets,
\cite{Savcheva2009} also find that EARs in the polar coronal holes statistically
follow Hale's polarity rule. Moreover, some time between early 2007 and late
2008 the orientation of these EARs changed from the one corresponding to cycle
24 to that of cycle 25. Early signs of the upcoming cycle are thus apparently
observed at high latitudes already around the start of the previous cycle.

It should be noted that detailed analyses of Hinode vector magnetograms of the
polar region (e.g. \cite{Shiota2012}) do not show features that might be
obviously identified with ephemeral active regions. As discussed above, magnetic
patches with fluxes above $10^{18}$ Mx are unipolar, while smaller flux elements
do not show any significant dependence on cycle, contrary to what is expected
for EAR. The solution of this puzzle is not known.

\section{Analysis of coronal green line emission data}

Emission from the solar corona is well known to be highly inhomogeneous,
depending on magnetic topology and field strength. Measurements of the intensity
of the coronal green line of Fe XIV at 5303 {\AA} slightly above the solar limb,
as a function of position angle have been made on a regular basis for a period
far exceeding the availability  and life time of space observatories; as a
consequence, long term variations in the corona covering several solar cycles
are most commonly analyzed using green line emission data.

A homogeneous data base of green line emission intensities has been compiled by
\cite{Minarovjech2011} and recently extended by cross calibration with SoHO EIT
data (\citen{Dorotovic2014}). Our investigation is based on this extended
homogeneous data set, available in a public data base.\footnote{\tt
http://www.suh.sk/online-data/modifikovany-homogenny-rad}

In this paper we focus on the RTTP due to its presumed link with the buildup of
the poloidal field. For a quantitative study of the RTTP time-latitude diagrams
(``butterfly diagrams'') of the green line emission need to be constructed with
sufficient contrast and detail at higher latitudes.

\begin{figure}
\epsfig{figure=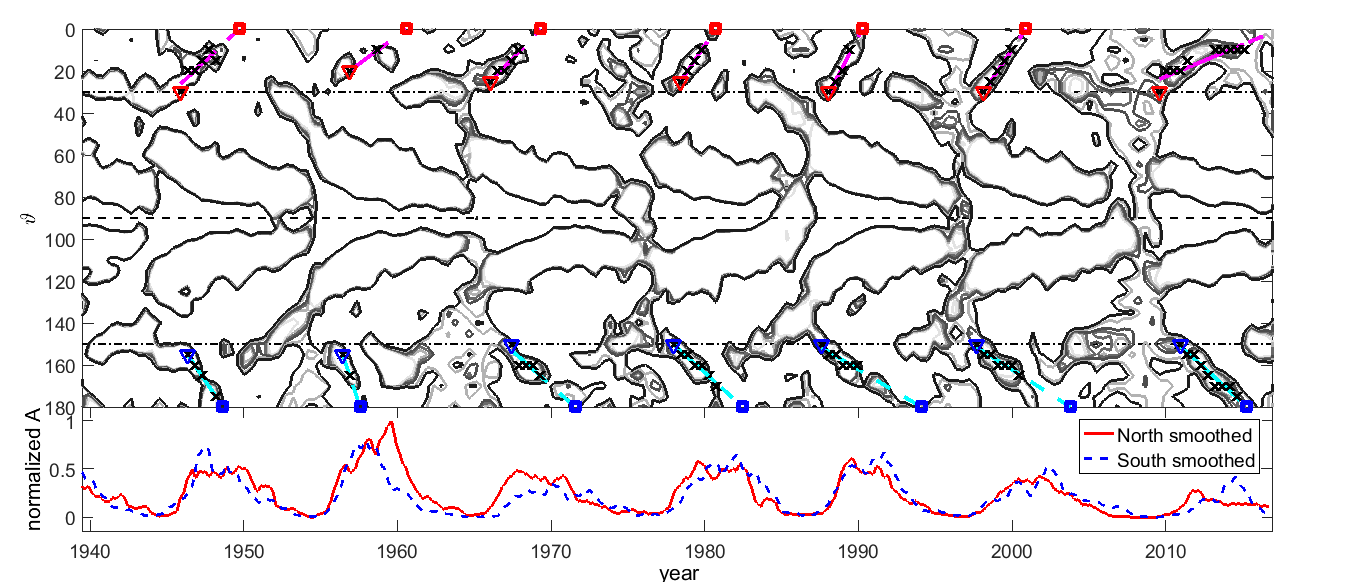,width=\textwidth}
\caption{11-month running averages of the coronal green line intensity after
applying unsharp masking with a 25$^\circ$ window in latitude (top panel). Crosses mark the
position of the RTTP, triangle the beginning, square the end of RTTP. Dashed lines are linear fits to these points. 
The bottom panel shows the smoothed, normalized sunspot area for each hemisphere (\tt http://solarscience.msfc.nasa.gov/greenwch.shtml).}
\label{fig:bfly}
\end{figure}

As high latitude structures in the corona are quite faint compared to the
low-latitude emission, simply plotting the detected intensity is not a suitable
method for the study of the polar coronal structures. Alternative methods used
by various authors for the visualization of these structures include plotting
the count of local maxima in the positional angle distribution of the emission
(e.g.\citen{Tappin+Altrock}); plotting the standard deviation of the intensity
relative to the mean value over a longer period (\citen{Leroy+Noens}); and
applying an  unsharp masking procedure to the butterfly diagram of the raw
intensity (\citen{Robbrecht}). Experimenting with all three methods 
we found
that the RTTP manifests itself most clearly in plots made with the unsharp
masking method, applying a masking window of width 25 degrees in latitude on the
11-month running mean of the raw intensity data, {\refrespb and then
subtracting this masked (blurred) image from the original, to enhance
fine details}.
The result is shown by contour
lines in Fig.~\ref{fig:bfly}. In what follows, we will present results based on
the data processed in this way.

In order to separate the polar component for quantitative analysis, the strongest 
positions above $60^{\circ}$ latitudes were identified. As starting point of RTTP 
component (crosses in Fig. \ref{fig:bfly}) we chose the epoch 
after which the pattern starts to approach the pole (see triangles in Fig. \ref{fig:bfly}). 
For each solar cycle and each hemisphere a linear fit was applied to these points to 
determine the end and tilt of the RTTP.

\begin{figure}
\center
\epsfig{figure=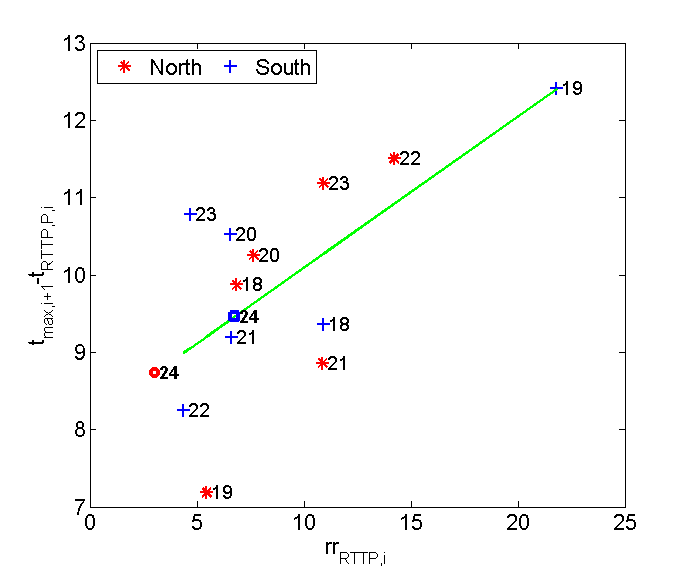,width=0.65\textwidth}
\caption{Rate of the RTTP $rr_{\mathrm{\scri RTTP}}$ (in degree/year)
against time (in years)  from the end of the RTTP $t_{\mathrm{\scri
RTTP},i}$ to the next solar maximum $R_{\mathrm{\scri max},i+1}$ for
individual cycles. The  correlation coefficient is $0.66$. The green
line is a linear fit to the data. Circle and square mark the
projections of the rate of the RTTP measured in cycle 24 (N and S
hemispheres) on this fit.}
\label{fig:predict}
\end{figure}

\section{Results}

The start and end dates of the RTTP as well as its tilt were plotted against a
variety of solar cycle parameters including amplitude, rise rate or time to
maximum. The best correlation
{\refresp ($r=0.66$)} 
is found in the case shown in Fig.~\ref{fig:predict}.

{\refresp 
As some of the RTTP's marked out in Fig.~\ref{fig:bfly} are rather
uncertainly defined, we also attempted a weighted fit to the data points in
Fig.~\ref{fig:predict}, the weights being the number of latitude grid points for
which RTTP positions (local maxima) were detected (crosses in
Fig.~\ref{fig:bfly}). The resulting fit, however, was found to
be almost indistinguishable to the one in Fig.~\ref{fig:predict}.
}

This correlation found between the rise rate of the RTTP and the time delay from
the ending of RTTP to the maximum of the following cycle may be used to  attempt
to predict the time of the maximum of cycle 25. 
{\refresp 
The fit shown in Fig.~\ref{fig:predict} may be written as
\begin{equation}
t_{\mathrm{max,}i+1} - t_{\mathrm{RTTP,}i} = a_1 \mathrm{rr}_{\mathrm{RTTP,}i} 
+ a_2
\label{eq:rttpfit}
\end{equation}
where $t_{\mathrm{max,}i}$ and $t_{\mathrm{RTTP,}i}$ are the epochs 
of the  maximum and of the end of the RTTP in cycle $i$; rr$_{\mathrm{RTTP,}i}$
is the rise rate of the RTTP in cycle $i$; while the $a_i$'s are
parameters of the fit.

Inserting into equation (\ref{eq:rttpfit}) the rate of the RTTP measured in
cycle 24 in the Southern hemisphere (where it is better defined in this cycle)
yields a time delay of 9.47 years from the time when the RTTP reached the poles
to the next maximum. It follows that the maximum of cycle 25 is expected to
occur in October 2024.
}

\begin{figure}
\epsfig{figure=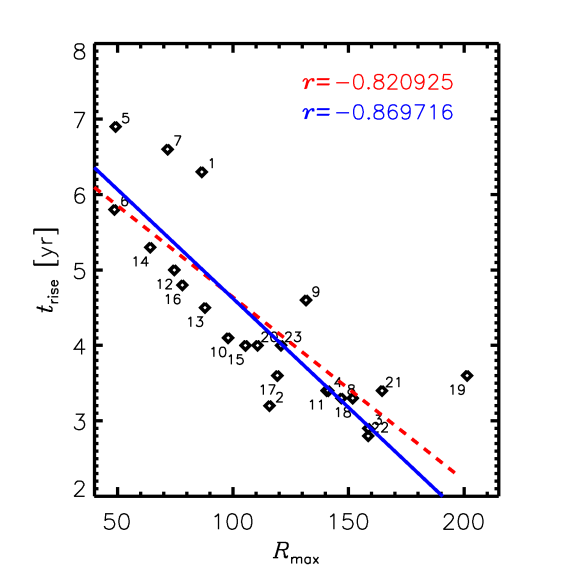,height=0.5\textwidth}
\epsfig{figure=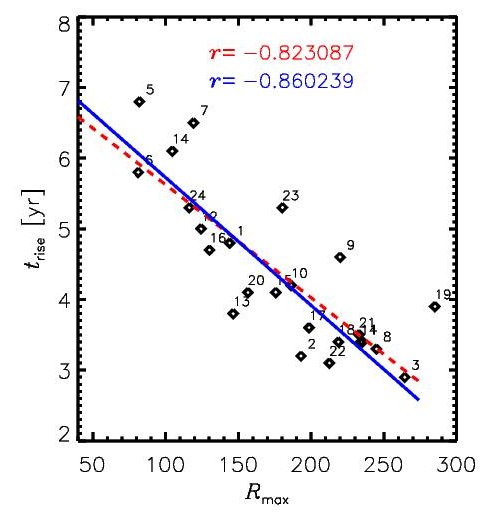,height=0.47\textwidth}
\caption{Correlation between cycle rise time $t_{\mathrm{\scri rise}}$
vs maximum amplitude $R_{\mathrm{\scri max}}$ (Waldmeier effect) with
unrevised (left) and revised (right) smoothed sunspot numbers. (Red
dashed: fit to all data points; blue solid: cycle 19 treated as
outlier.)}
\label{fig:Waldmoldnew}
\end{figure}

\begin{figure}
\epsfig{figure=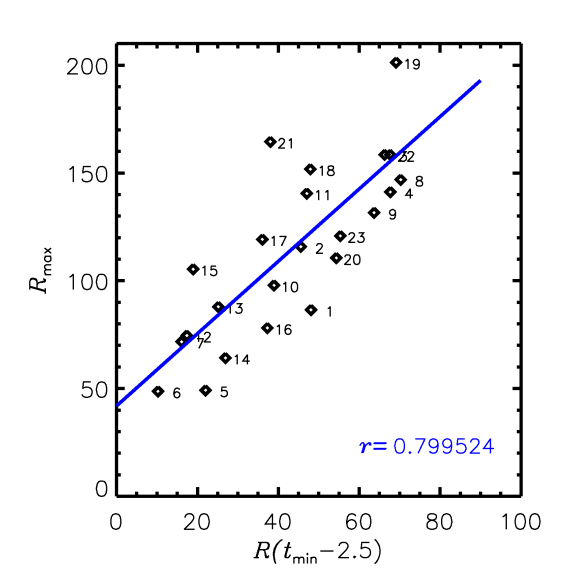,height=0.5\textwidth}
\epsfig{figure=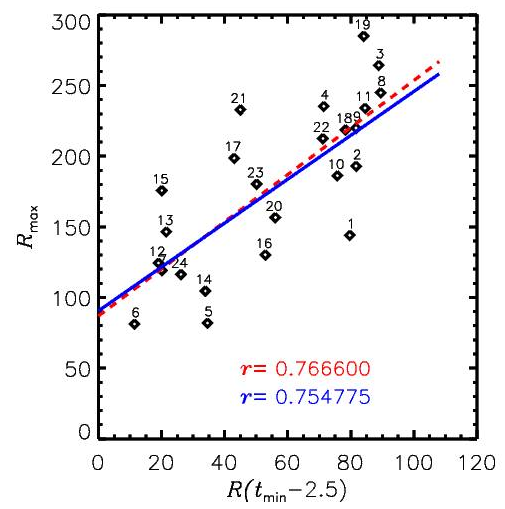,height=0.47\textwidth}
\caption{Correlation between maximum cycle amplitude $R_{\mathrm{\scri
max}}$ and sunspot number value 2.5 years before the previous minimum 
$R(t_\mathrm{\scri min}-2.5)$ for unrevised (left) and revised (right)
smoothed sunspot numbers.
(Red dashed: fit to all data points; blue solid: cycle 19 treated as outlier.)}
\label{fig:CScheff}
\end{figure}

\section{Conclusion}

Based on an analysis of the rush-to-the-poles phenomenon in coronal green line
data we concluded that the rate of the RTTP is correlated with the time delay to
to the maximum of the next solar cycle counted from the time when the RTTP
reaches the pole. On the basis of this result we predict the next solar maximum to
to occur in 
{\refresp 
October 2024.

In addition to the correlation found above, equation (\ref{eq:rttpfit}), two
other, quite robust correlations exist between parameters of subsequent cycles
(cf. Figs \ref{fig:Waldmoldnew} and \ref{fig:CScheff}).
One of these is the Waldmeier effect {\refrespc (\citen{Waldmeier:effect})}
\begin{equation}
t_{\mathrm{max,}i+1} - t_{\mathrm{min,}i+1} = a_3 R_{\mathrm{max,}i+1} + a_4
\label{eq:Waldm}
\end{equation}
where  $t_{\mathrm{min,}i}$ is the epoch of
the minimum starting cycle $i$ while $R_{\mathrm{max,}i}$ is the maximum value
of the smoothed sunpot number $R$ in the cycle. The other known correlation is
the ``minimax3'' rule discussed in \cite{Petrovay:LRSP} and originally due to
\cite{CSch2007}:
\begin{equation}
R_{\mathrm{max,}i+1} = a_5 R(t_{\mathrm{min,}i+1}-2.5) + a_6
\label{eq:minimax3}
\end{equation}
where $R(t)$ is the smoothed sunspot number series, known from observations and
here evaluated 2.5 years before the minimum.

Once we determined $t_{\mathrm{max,}i+1}$ from equation (\ref{eq:rttpfit}),
a numerical solution of equations (\ref{eq:Waldm}) and (\ref{eq:minimax3})
yields $R_{\mathrm{max,}i+1}=129$ and $t_{\mathrm{min,}i+1}=2019.4$. 

In summary, based on the apparent correlation between the rate ofthe RTTP and
the time delay to the next maximum and the known internal regularities of the
sunspot number series we predict that, following a minimum in 2019, cycle 25
will most likely peak in late 2024 at an amplitude of about 130. This slightly
exceeds the amplitude of cycle 24 but it would still make cycle 25 a fairly weak
cycle.
}

\section*{Acknowledgement}
{\refrespc This paper includes material first presented at the ISSI/VarSITI-SEE 
forum on ``Expected evolution of solar activity in the following
decades'', held in March 2016 in Bern.
The hospitality of the organizers is gratefully acknowledged.}
This research was partially funded by the European Union's Horizon 2020 
research and innovation programme under grant agreement No. 739500.
SILSO sunspot number data used in this study are provided by the Royal
Observatory of Belgium,  Brussels.
Wilcox Solar Observatory data used in this study was obtained via the web site
\url{http://wso.stanford.edu/}, courtesy of J.T.~Hoeksema.
The Wilcox Solar Observatory is currently supported by NASA.


\bibliography{greenlpred}

\begin{thebibliography}{32}
\expandafter\ifx\csname natexlab\endcsname\relax\def\natexlab#1{#1}\fi
\providecommand{\url}[1]{\texttt{#1}}
\providecommand{\href}[2]{#2}
\providecommand{\path}[1]{#1}
\providecommand{\DOIprefix}{doi:}
\providecommand{\ArXivprefix}{arXiv:}
\providecommand{\URLprefix}{URL: }
\providecommand{\Pubmedprefix}{pmid:}
\providecommand{\doi}[1]{\href{http://dx.doi.org/#1}{\path{#1}}}
\providecommand{\Pubmed}[1]{\href{pmid:#1}{\path{#1}}}
\providecommand{\bibinfo}[2]{#2}
\ifx\xfnm\relax \def\xfnm[#1]{\unskip,\space#1}\fi
\bibitem[{{Blanco Rodr{\'{\i}}guez} and {Kneer}(2010)}]{Blanco2010}
\bibinfo{author}{{Blanco Rodr{\'{\i}}guez}, J.}, \bibinfo{author}{{Kneer}, F.},
  \bibinfo{year}{2010}.
\newblock \bibinfo{title}{{Faculae at the poles of the Sun revisited: infrared
  observations}}.
\newblock \bibinfo{journal}{\aap} \bibinfo{volume}{509}, \bibinfo{pages}{A92}.
\newblock \DOIprefix\doi{10.1051/0004-6361/200811111}.
\bibitem[{Cameron and {Sch{\"u}ssler}(2007)}]{CSch2007}
\bibinfo{author}{Cameron, R.}, \bibinfo{author}{{Sch{\"u}ssler}, M.},
  \bibinfo{year}{2007}.
\newblock \bibinfo{title}{Solar cycle prediction using precursors and flux
  transport models}.
\newblock \bibinfo{journal}{Astrophys. J.} \bibinfo{volume}{659},
  \bibinfo{pages}{801--811}.
\newblock \DOIprefix\doi{10.1086/512049},
  \href{http://arxiv.org/abs/arXiv:astro-ph/0612693}{\tt
  arXiv:arXiv:astro-ph/0612693}.
\bibitem[{{Dorotovi{\v c}} et~al.(2014){Dorotovi{\v c}}, {Minarovjech},
  {Lorenc} and {Rybansk{\'y}}}]{Dorotovic2014}
\bibinfo{author}{{Dorotovi{\v c}}, I.}, \bibinfo{author}{{Minarovjech}, M.},
  \bibinfo{author}{{Lorenc}, M.}, \bibinfo{author}{{Rybansk{\'y}}, M.},
  \bibinfo{year}{2014}.
\newblock \bibinfo{title}{{Modified Homogeneous Data Set of Coronal
  Intensities}}.
\newblock \bibinfo{journal}{\solphys} \bibinfo{volume}{289},
  \bibinfo{pages}{2697--2703}.
\newblock \DOIprefix\doi{10.1007/s11207-014-0501-2}.
\bibitem[{{Hagenaar}(2001)}]{Hagenaar2001}
\bibinfo{author}{{Hagenaar}, H.J.}, \bibinfo{year}{2001}.
\newblock \bibinfo{title}{{Ephemeral Regions on a Sequence of Full-Disk
  Michelson Doppler Imager Magnetograms}}.
\newblock \bibinfo{journal}{\apj} \bibinfo{volume}{555},
  \bibinfo{pages}{448--461}.
\newblock \DOIprefix\doi{10.1086/321448}.
\bibitem[{{Howe}(2009)}]{Howe2009}
\bibinfo{author}{{Howe}, R.}, \bibinfo{year}{2009}.
\newblock \bibinfo{title}{{Solar Interior Rotation and its Variation}}.
\newblock \bibinfo{journal}{Living Reviews in Solar Physics}
  \bibinfo{volume}{6}, \bibinfo{pages}{1}.
\newblock \DOIprefix\doi{10.12942/lrsp-2009-1},
  \href{http://arxiv.org/abs/0902.2406}{\tt arXiv:0902.2406}.
\bibitem[{{Howe} et~al.(2013){Howe}, {Christensen-Dalsgaard}, {Hill}, {Komm},
  {Larson}, {Rempel}, {Schou} and {Thompson}}]{Howe2013}
\bibinfo{author}{{Howe}, R.}, \bibinfo{author}{{Christensen-Dalsgaard}, J.},
  \bibinfo{author}{{Hill}, F.}, \bibinfo{author}{{Komm}, R.},
  \bibinfo{author}{{Larson}, T.P.}, \bibinfo{author}{{Rempel}, M.},
  \bibinfo{author}{{Schou}, J.}, \bibinfo{author}{{Thompson}, M.J.},
  \bibinfo{year}{2013}.
\newblock \bibinfo{title}{{The High-latitude Branch of the Solar Torsional
  Oscillation in the Rising Phase of Cycle 24}}.
\newblock \bibinfo{journal}{\apjl} \bibinfo{volume}{767}, \bibinfo{pages}{L20}.
\newblock \DOIprefix\doi{10.1088/2041-8205/767/1/L20}.
\bibitem[{{Howe} et~al.(2006){Howe}, {Komm}, {Hill}, {Ulrich}, {Haber},
  {Hindman}, {Schou} and {Thompson}}]{Howe2006}
\bibinfo{author}{{Howe}, R.}, \bibinfo{author}{{Komm}, R.},
  \bibinfo{author}{{Hill}, F.}, \bibinfo{author}{{Ulrich}, R.},
  \bibinfo{author}{{Haber}, D.A.}, \bibinfo{author}{{Hindman}, B.W.},
  \bibinfo{author}{{Schou}, J.}, \bibinfo{author}{{Thompson}, M.J.},
  \bibinfo{year}{2006}.
\newblock \bibinfo{title}{{Large-Scale Zonal Flows Near the Solar Surface}}.
\newblock \bibinfo{journal}{\solphys} \bibinfo{volume}{235},
  \bibinfo{pages}{1--15}.
\newblock \DOIprefix\doi{10.1007/s11207-006-0117-2}.
\bibitem[{{Kaithakkal} et~al.(2013){Kaithakkal}, {Suematsu}, {Kubo}, {Shiota}
  and {Tsuneta}}]{Kaithakkal2013}
\bibinfo{author}{{Kaithakkal}, A.J.}, \bibinfo{author}{{Suematsu}, Y.},
  \bibinfo{author}{{Kubo}, M.}, \bibinfo{author}{{Shiota}, D.},
  \bibinfo{author}{{Tsuneta}, S.}, \bibinfo{year}{2013}.
\newblock \bibinfo{title}{{The Association of Polar Faculae with Polar Magnetic
  Patches Examined with Hinode Observations}}.
\newblock \bibinfo{journal}{\apj} \bibinfo{volume}{776}, \bibinfo{pages}{122}.
\newblock \DOIprefix\doi{10.1088/0004-637X/776/2/122},
  \href{http://arxiv.org/abs/1311.0980}{\tt arXiv:1311.0980}.
\bibitem[{{Leroy} and {Noens}(1983)}]{Leroy+Noens}
\bibinfo{author}{{Leroy}, J.L.}, \bibinfo{author}{{Noens}, J.C.},
  \bibinfo{year}{1983}.
\newblock \bibinfo{title}{{Does the solar activity cycle extend over more than
  an 11-year period?}}
\newblock \bibinfo{journal}{\aap} \bibinfo{volume}{120}, \bibinfo{pages}{L1}.
\bibitem[{{Martin} and {Harvey}(1979)}]{Martin+Harvey79}
\bibinfo{author}{{Martin}, S.F.}, \bibinfo{author}{{Harvey}, K.H.},
  \bibinfo{year}{1979}.
\newblock \bibinfo{title}{{Ephemeral active regions during solar minimum}}.
\newblock \bibinfo{journal}{\solphys} \bibinfo{volume}{64},
  \bibinfo{pages}{93--108}.
\newblock \DOIprefix\doi{10.1007/BF00151118}.
\bibitem[{{McIntosh} and {Leamon}(2017)}]{McIntosh2017}
\bibinfo{author}{{McIntosh}, S.W.}, \bibinfo{author}{{Leamon}, R.J.},
  \bibinfo{year}{2017}.
\newblock \bibinfo{title}{{Deciphering Solar Magnetic Activity: Spotting Solar
  Cycle 25}}.
\newblock \bibinfo{journal}{ArXiv e-prints}
  \href{http://arxiv.org/abs/1702.04414}{\tt arXiv:1702.04414}.
\bibitem[{{McIntosh} et~al.(2014){McIntosh}, {Wang}, {Leamon}, {Davey}, {Howe},
  {Krista}, {Malanushenko}, {Markel}, {Cirtain}, {Gurman}, {Pesnell} and
  {Thompson}}]{McIntosh2014}
\bibinfo{author}{{McIntosh}, S.W.}, \bibinfo{author}{{Wang}, X.},
  \bibinfo{author}{{Leamon}, R.J.}, \bibinfo{author}{{Davey}, A.R.},
  \bibinfo{author}{{Howe}, R.}, \bibinfo{author}{{Krista}, L.D.},
  \bibinfo{author}{{Malanushenko}, A.V.}, \bibinfo{author}{{Markel}, R.S.},
  \bibinfo{author}{{Cirtain}, J.W.}, \bibinfo{author}{{Gurman}, J.B.},
  \bibinfo{author}{{Pesnell}, W.D.}, \bibinfo{author}{{Thompson}, M.J.},
  \bibinfo{year}{2014}.
\newblock \bibinfo{title}{{Deciphering Solar Magnetic Activity. I. On the
  Relationship between the Sunspot Cycle and the Evolution of Small Magnetic
  Features}}.
\newblock \bibinfo{journal}{\apj} \bibinfo{volume}{792}, \bibinfo{pages}{12}.
\newblock \DOIprefix\doi{10.1088/0004-637X/792/1/12},
  \href{http://arxiv.org/abs/1403.3071}{\tt arXiv:1403.3071}.
\bibitem[{{Minarovjech} et~al.(2011){Minarovjech}, {Ru{\v s}in} and
  {Saniga}}]{Minarovjech2011}
\bibinfo{author}{{Minarovjech}, M.}, \bibinfo{author}{{Ru{\v s}in}, V.},
  \bibinfo{author}{{Saniga}, M.}, \bibinfo{year}{2011}.
\newblock \bibinfo{title}{{The green corona database and the coronal index of
  solar activity}}.
\newblock \bibinfo{journal}{Contributions of the Astronomical Observatory
  Skalnate Pleso} \bibinfo{volume}{41}, \bibinfo{pages}{137--141}.
\bibitem[{{Mu{\~n}oz-Jaramillo} et~al.(2012){Mu{\~n}oz-Jaramillo}, {Sheeley},
  {Zhang} and {DeLuca}}]{Munozjara2012}
\bibinfo{author}{{Mu{\~n}oz-Jaramillo}, A.}, \bibinfo{author}{{Sheeley}, N.R.},
  \bibinfo{author}{{Zhang}, J.}, \bibinfo{author}{{DeLuca}, E.E.},
  \bibinfo{year}{2012}.
\newblock \bibinfo{title}{{Calibrating 100 Years of Polar Faculae Measurements:
  Implications for the Evolution of the Heliospheric Magnetic Field}}.
\newblock \bibinfo{journal}{\apj} \bibinfo{volume}{753}, \bibinfo{pages}{146}.
\newblock \DOIprefix\doi{10.1088/0004-637X/753/2/146},
  \href{http://arxiv.org/abs/1303.0345}{\tt arXiv:1303.0345}.
\bibitem[{{Petrie}(2015)}]{PetrieLRSP}
\bibinfo{author}{{Petrie}, G.J.D.}, \bibinfo{year}{2015}.
\newblock \bibinfo{title}{{Solar Magnetism in the Polar Regions}}.
\newblock \bibinfo{journal}{Living Reviews in Solar Physics}
  \bibinfo{volume}{12}, \bibinfo{pages}{5}.
\newblock \DOIprefix\doi{10.1007/lrsp-2015-5}.
\bibitem[{{Petrie} et~al.(2014){Petrie}, {Petrovay} and {Schatten}}]{Petrie+}
\bibinfo{author}{{Petrie}, G.J.D.}, \bibinfo{author}{{Petrovay}, K.},
  \bibinfo{author}{{Schatten}, K.}, \bibinfo{year}{2014}.
\newblock \bibinfo{title}{{Solar Polar Fields and the 22-Year Activity Cycle:
  Observations and Models}}.
\newblock \bibinfo{journal}{\ssr} \bibinfo{volume}{186},
  \bibinfo{pages}{325--357}.
\newblock \DOIprefix\doi{10.1007/s11214-014-0064-4}.
\bibitem[{{Petrovay}(2010)}]{Petrovay:LRSP}
\bibinfo{author}{{Petrovay}, K.}, \bibinfo{year}{2010}.
\newblock \bibinfo{title}{{Solar Cycle Prediction}}.
\newblock \bibinfo{journal}{Living Reviews in Solar Physics}
  \bibinfo{volume}{7}, \bibinfo{pages}{6}.
\newblock \DOIprefix\doi{10.12942/lrsp-2010-6},
  \href{http://arxiv.org/abs/1012.5513}{\tt arXiv:1012.5513}.
\bibitem[{Petrovay and Forg{\'a}cs-Dajka(2002)}]{PetrovayFDE2002}
\bibinfo{author}{Petrovay, K.}, \bibinfo{author}{Forg{\'a}cs-Dajka, E.},
  \bibinfo{year}{2002}.
\newblock \bibinfo{title}{{The Role of Active Regions in the Generation of
  Torsional Oscillations}}.
\newblock \bibinfo{journal}{\solphys} \bibinfo{volume}{205},
  \bibinfo{pages}{39--52}.
\bibitem[{{Priyal} et~al.(2014){Priyal}, {Banerjee}, {Karak},
  {Mu{\~n}oz-Jaramillo}, {Ravindra}, {Choudhuri} and {Singh}}]{Priyal2014}
\bibinfo{author}{{Priyal}, M.}, \bibinfo{author}{{Banerjee}, D.},
  \bibinfo{author}{{Karak}, B.B.}, \bibinfo{author}{{Mu{\~n}oz-Jaramillo}, A.},
  \bibinfo{author}{{Ravindra}, B.}, \bibinfo{author}{{Choudhuri}, A.R.},
  \bibinfo{author}{{Singh}, J.}, \bibinfo{year}{2014}.
\newblock \bibinfo{title}{{Polar Network Index as a Magnetic Proxy for the
  Solar Cycle Studies}}.
\newblock \bibinfo{journal}{\apjl} \bibinfo{volume}{793}, \bibinfo{pages}{L4}.
\newblock \DOIprefix\doi{10.1088/2041-8205/793/1/L4},
  \href{http://arxiv.org/abs/1407.4944}{\tt arXiv:1407.4944}.
\bibitem[{{Robbrecht} et~al.(2010){Robbrecht}, {Wang}, {Sheeley} and
  {Rich}}]{Robbrecht}
\bibinfo{author}{{Robbrecht}, E.}, \bibinfo{author}{{Wang}, Y.M.},
  \bibinfo{author}{{Sheeley}, Jr., N.R.}, \bibinfo{author}{{Rich}, N.B.},
  \bibinfo{year}{2010}.
\newblock \bibinfo{title}{{On the ''Extended'' Solar Cycle in Coronal
  Emission}}.
\newblock \bibinfo{journal}{\apj} \bibinfo{volume}{716},
  \bibinfo{pages}{693--700}.
\newblock \DOIprefix\doi{10.1088/0004-637X/716/1/693}.
\bibitem[{{Savcheva} et~al.(2009){Savcheva}, {Cirtain}, {DeLuca} and
  {Golub}}]{Savcheva2009}
\bibinfo{author}{{Savcheva}, A.}, \bibinfo{author}{{Cirtain}, J.W.},
  \bibinfo{author}{{DeLuca}, E.E.}, \bibinfo{author}{{Golub}, L.},
  \bibinfo{year}{2009}.
\newblock \bibinfo{title}{{Does a Polar Coronal Hole's Flux Emergence Follow a
  Hale-Like Law?}}
\newblock \bibinfo{journal}{\apjl} \bibinfo{volume}{702},
  \bibinfo{pages}{L32--L36}.
\newblock \DOIprefix\doi{10.1088/0004-637X/702/1/L32}.
\bibitem[{{Scherrer} et~al.(1977){Scherrer}, {Wilcox}, {Svalgaard}, {Duvall},
  {Dittmer} and {Gustafson}}]{WSO}
\bibinfo{author}{{Scherrer}, P.H.}, \bibinfo{author}{{Wilcox}, J.M.},
  \bibinfo{author}{{Svalgaard}, L.}, \bibinfo{author}{{Duvall}, Jr., T.L.},
  \bibinfo{author}{{Dittmer}, P.H.}, \bibinfo{author}{{Gustafson}, E.K.},
  \bibinfo{year}{1977}.
\newblock \bibinfo{title}{{The mean magnetic field of the sun - Observations at
  Stanford}}.
\newblock \bibinfo{journal}{\solphys} \bibinfo{volume}{54},
  \bibinfo{pages}{353--361}.
\newblock \DOIprefix\doi{10.1007/BF00159925}.
\bibitem[{{Shiota} et~al.(2012){Shiota}, {Tsuneta}, {Shimojo}, {Sako}, {Orozco
  Su{\'a}rez} and {Ishikawa}}]{Shiota2012}
\bibinfo{author}{{Shiota}, D.}, \bibinfo{author}{{Tsuneta}, S.},
  \bibinfo{author}{{Shimojo}, M.}, \bibinfo{author}{{Sako}, N.},
  \bibinfo{author}{{Orozco Su{\'a}rez}, D.}, \bibinfo{author}{{Ishikawa}, R.},
  \bibinfo{year}{2012}.
\newblock \bibinfo{title}{{Polar Field Reversal Observations with Hinode}}.
\newblock \bibinfo{journal}{\apj} \bibinfo{volume}{753}, \bibinfo{pages}{157}.
\newblock \DOIprefix\doi{10.1088/0004-637X/753/2/157},
  \href{http://arxiv.org/abs/1205.2154}{\tt arXiv:1205.2154}.
\bibitem[{{SILSO WDC}(2017)}]{SILSO}
\bibinfo{author}{{SILSO WDC}}, \bibinfo{year}{2017}.
\newblock \bibinfo{title}{{The International Sunspot Number}}.
\newblock \bibinfo{journal}{International Sunspot Number Monthly Bulletin and
  online catalogue} .
\bibitem[{{Spruit}(2003)}]{Spruit}
\bibinfo{author}{{Spruit}, H.C.}, \bibinfo{year}{2003}.
\newblock \bibinfo{title}{{Origin of the torsional oscillation pattern of solar
  rotation}}.
\newblock \bibinfo{journal}{\solphys} \bibinfo{volume}{213},
  \bibinfo{pages}{1--21}.
\newblock \DOIprefix\doi{10.1023/A:1023202605379},
  \href{http://arxiv.org/abs/astro-ph/0209146}{\tt arXiv:astro-ph/0209146}.
\bibitem[{{Sun} et~al.(2015){Sun}, {Hoeksema}, {Liu} and {Zhao}}]{Sun2015}
\bibinfo{author}{{Sun}, X.}, \bibinfo{author}{{Hoeksema}, J.T.},
  \bibinfo{author}{{Liu}, Y.}, \bibinfo{author}{{Zhao}, J.},
  \bibinfo{year}{2015}.
\newblock \bibinfo{title}{{On Polar Magnetic Field Reversal and Surface Flux
  Transport During Solar Cycle 24}}.
\newblock \bibinfo{journal}{\apj} \bibinfo{volume}{798}, \bibinfo{pages}{114}.
\newblock \DOIprefix\doi{10.1088/0004-637X/798/2/114},
  \href{http://arxiv.org/abs/1410.8867}{\tt arXiv:1410.8867}.
\bibitem[{{Tappin} and {Altrock}(2013)}]{Tappin+Altrock}
\bibinfo{author}{{Tappin}, S.J.}, \bibinfo{author}{{Altrock}, R.C.},
  \bibinfo{year}{2013}.
\newblock \bibinfo{title}{{The Extended Solar Cycle Tracked High into the
  Corona}}.
\newblock \bibinfo{journal}{\solphys} \bibinfo{volume}{282},
  \bibinfo{pages}{249--261}.
\newblock \DOIprefix\doi{10.1007/s11207-012-0133-3},
  \href{http://arxiv.org/abs/1209.2969}{\tt arXiv:1209.2969}.
\bibitem[{Tlatov et~al.(2010)Tlatov, {Vasil'eva} and {Pevtsov}}]{Tlatov2010}
\bibinfo{author}{Tlatov, A.}, \bibinfo{author}{{Vasil'eva}, V.},
  \bibinfo{author}{{Pevtsov}, A.}, \bibinfo{year}{2010}.
\newblock \bibinfo{title}{Distribution of magnetic bipoles on the sun over
  three solar cycles}.
\newblock \bibinfo{journal}{Astrophys. J.} \bibinfo{volume}{717},
  \bibinfo{pages}{357--362}.
\newblock \DOIprefix\doi{10.1088/0004-637X/717/1/357}.
\bibitem[{{Tlatov} et~al.(2015){Tlatov}, {Dormidontov}, {Kirpichev},
  {Pashchenko}, {Shramko}, {Peshcherov}, {Grigoryev}, {Demidov} and
  {Svidskii}}]{Tlatov2015}
\bibinfo{author}{{Tlatov}, A.G.}, \bibinfo{author}{{Dormidontov}, D.V.},
  \bibinfo{author}{{Kirpichev}, R.V.}, \bibinfo{author}{{Pashchenko}, M.P.},
  \bibinfo{author}{{Shramko}, A.D.}, \bibinfo{author}{{Peshcherov}, V.S.},
  \bibinfo{author}{{Grigoryev}, V.M.}, \bibinfo{author}{{Demidov}, M.L.},
  \bibinfo{author}{{Svidskii}, P.M.}, \bibinfo{year}{2015}.
\newblock \bibinfo{title}{{Study of some characteristics of large-scale solar
  magnetic fields during the global field polarity reversal according to
  observations at the telescope-magnetograph Kislovodsk Observatory}}.
\newblock \bibinfo{journal}{Geomagnetism and Aeronomy} \bibinfo{volume}{55},
  \bibinfo{pages}{969--975}.
\newblock \DOIprefix\doi{10.1134/S0016793215070257}.
\bibitem[{{Ulrich} et~al.(1988){Ulrich}, {Boyden}, {Webster}, {Padilla} and
  {Snodgrass}}]{Ulrich1988}
\bibinfo{author}{{Ulrich}, R.K.}, \bibinfo{author}{{Boyden}, J.E.},
  \bibinfo{author}{{Webster}, L.}, \bibinfo{author}{{Padilla}, S.P.},
  \bibinfo{author}{{Snodgrass}, H.B.}, \bibinfo{year}{1988}.
\newblock \bibinfo{title}{{Solar rotation measurements at Mount Wilson. V -
  Reanalysis of 21 years of data}}.
\newblock \bibinfo{journal}{\solphys} \bibinfo{volume}{117},
  \bibinfo{pages}{291--328}.
\newblock \DOIprefix\doi{10.1007/BF00147250}.
\bibitem[{Waldmeier(1935)}]{Waldmeier:effect}
\bibinfo{author}{Waldmeier, M.}, \bibinfo{year}{1935}.
\newblock \bibinfo{journal}{Astr. Mitt. Z{\"u}rich} \bibinfo{volume}{14 (133)},
  \bibinfo{pages}{105--130}.
\bibitem[{Wilson et~al.(1988)Wilson, {Altrock}, {Harvey}, {Martin} and
  {Snodgrass}}]{Wilson1988}
\bibinfo{author}{Wilson, P.}, \bibinfo{author}{{Altrock}, R.},
  \bibinfo{author}{{Harvey}, K.}, \bibinfo{author}{{Martin}, S.},
  \bibinfo{author}{{Snodgrass}, H.}, \bibinfo{year}{1988}.
\newblock \bibinfo{title}{The extended solar activity cycle}.
\newblock \bibinfo{journal}{Nature} \bibinfo{volume}{333},
  \bibinfo{pages}{748--+}.
\newblock \DOIprefix\doi{10.1038/333748a0}.

\end{thebibliography}

\end{document}